\documentclass[preprint2]{aastex}

\usepackage{hyperref}
\usepackage{graphicx}

\begin{document}

%% LaTeX will automatically break titles if they run longer than
%% one line. However, you may use \\ to force a line break if you
%% desire.

\title{Detection of a $50\arcdeg$-long Trailing Tidal Tail for the
  Globular Cluster M5}

\author{Carl J. Grillmair}
\affil{IPAC, California Institute of Technology, Pasadena, CA 91125}
\email{carl@ipac.caltech.edu}

\begin{abstract}

Using photometry and proper motions from Gaia Data Release 2, we
detect a $50\arcdeg$-long stream of about 70 stars extending westward
from the halo globular cluster M5. Based on the similarities in
distance, proper motions, inferred color-magnitude distribution, and
trajectory, we identify this stream as the trailing tidal tail of
M5. While the surface density of stars is very low ($\simeq 1.5$ stars
per square degree, or $\approx 35$ magnitudes per square arcsecond),
selecting only stars having proper motions consistent with the orbit
of the cluster yield a detection significance $\approx
10\sigma$. While we find a possible continuation of the stream to
$\approx 85\arcdeg$, increasing foreground contamination combined with
a greater predicted stream distance make it difficult to detect with
current data even if the stream continues unabated. The non-uniform
distribution of stars in the stream appears to be consistent with episodic
tidal stripping, with the most recently shed stars now trailing the
cluster by tens of degrees. We provide a table of the highest-ranked
candidate stream stars for ongoing and future spectroscopic surveys.

\end{abstract}

%% Keywords should appear after the \end{abstract} command. The uncommented
%% example has been keyed in ApJ style. See the instructions to authors
%% for the journal to which you are submitting your paper to determine
%% what keyword punctuation is appropriate.


\section{Introduction}

The inner Galactic halo is now known to be populated with dozens of
stellar debris streams \citep{grillmair2016, shipp2018, malhan2018b,
  ibata2019, palau2019}. Most of these streams are relatively
narrow, with physical widths on the order of 100 pc, and they are
consequently assumed to have been produced by globular clusters. Yet
we know of only three extant globular clusters with long ($>
10\arcdeg$) tidal tails, namely Pal 5 \citep{odenkirchen2003}, NGC
5466 \citep{grillmair2006, belokurov2006}, and M68
\citep{palau2019}. This is somewhat surprising, given that most
globular clusters studied to date show evidence for departures from
King profiles \citep{king1966} in their outskirts that are consistent
with the development of tidal tails \citep{grillmair1995, leon2000}.

With episodic stripping and consequent regular variations in stream
density, tidal tails extending from extant globular clusters may be
somewhat problematic for detecting signatures of dark matter
subhalos \citep{kupper2012}. On the other hand, by providing well-characterized
progenitors, such tidal tails will be useful for understanding the
detailed physics of tidal stripping, the accretion sequence of the
halo, and the shape of the Galactic potential \citep{bovy2016}.

With the second data release of the Gaia catalog we now have
considerably more information for sorting stars into
substructures. \citet{malhan2018a}, \citet{malhan2018b} and
\citet{ibata2019} have applied brute force orbital integration
techniques to discover many halo streams that would have been
impossible to detect with the purely photometric techniques of a few
years ago. Here we use an alternative, more directed method to detect
a long trailing tail behind the globular cluster M5. Section
\ref{analysis} describes our method. We discuss possible issues with
the association of the stream with M5 in Section \ref{discussion} and
make concluding remarks in Section \ref{conclusions}.

\section{Analysis} \label{analysis}

Our search technique makes use of the Gaia second data release
\citep{gaia2016, gaia2018a}. While the relatively bright limiting
magnitude of the catalog ($G \approx 20$) rather limits the numbers of
main sequence stars that can be sampled (compared with deeper
photometric catalogs like the Sloan Digital Sky Survey or Pan-STARRS) the
addition of high-quality proper motion measurements enables selection
criteria that can more than compensate in certain circumstances.

\citet{gaia2018b} used the Gaia catalog to measure proper motions for
most of the known Galactic globular clusters. Using these proper
motions along with previously published radial velocities, we used a
simple model of the Galaxy (\citet{allen1991}, updated using the
parameters of \citet{irrgang2013}) to compute orbits for clusters.
From these orbits we derived expectation profiles for proper motions
as a function of position on the sky. The predicted run of position,
distance, radial velocity, and proper motions for M5's orbit are shown
in Figure 1.

We then applied a modified form of the matched filter described by
\citet{rockosi2002} and \citet{grillmair2009} to the Gaia data.
Briefly, we used the Gaia-observed $G, G_{BP} - G_{RP}$ distribution
of stars in M5 to produce a color-magnitude locus extending from the
apparent tip of the red giant branch to the main sequence at $G
\approx 21$. This locus was then shifted up and down in $G$ to account
for possible changes in distance of stream stars. At each magnitude
shift/assumed distance, individual stars in the field were accorded
weights based on their minimum $G, G_{BP} - G_{RP}$ distances from the
CM locus. These weights were computed using the Gaia-provided
photometric uncertainties and assuming a Gaussian error
distribution. We used only stars with photometric uncertainties of
less than 0.5 mag and valid proper motion measurements. In this newly
modified form of the filter, these weights were then scaled again
using each stars's departure from the expected proper motion profiles
shown in Figure 1. The Gaia-provided proper motion errors were assumed
to be Gaussian and the corresponding weights computed as:

\begin{equation}
\label{mfilter}
w \propto \exp{\{-0.5 \left[\left({\mu_{\alpha} - \mu_{\alpha, pred}\over\sigma_{\mu_{\alpha}}}\right)^2 + \left({\mu_{\delta} - \mu_{\delta, pred})\over\sigma_{\mu_{\delta}}}\right)^2\right]\}}
\end{equation}

\noindent where $\mu_{pred}$ are the components of proper motion
predicted at each star's Right Ascension in Figure 1. We then summed
these weights by sky position to produce the weighted surface density
map shown in Figure 2.

Figure 2 shows a tenuous but significant stream of stars extending
$\approx 50\arcdeg$ to the northwest of M5. The stream is not aligned
with any features in Gaia's scan pattern, nor does it coincide with
any extended regions of particularly low reddening
\citep{schlegel1998}. The stream is visible only over a limited range
of assumed distance moduli, strengthening and subsiding with
increasing distance in a manner that is characteristic of purely
photometric uncertainties. While there is no obvious M5-like
color-magnitude sequence visible for purely proper motion-selected
stars within the area of the stream, using color-magnitude filters
that are either somewhat more metal-rich or more metal-poor than M5
yield a reduced maximum signal-to-noise ratio (SNR). Based on the
strength of the matched-filtered signal, the stream appears to be
located about 0.1 magnitudes closer over of much of its length than
the 7.5 kpc distance of M5 itself \citep{harris1996}. This
qualitatively agrees with Figure 1, which predicts a heliocentric
distance that drops slightly to 7.3 kpc at R.A. $\approx 217\arcdeg$
and then increases to nearly 15 kpc at R.A. $\approx 134\arcdeg$.

Over the region $190\arcdeg <$ R.A. $< 225\arcdeg$ the path of the stream can be modeled to within $0.3\arcdeg$ using:

\begin{equation}
\label{n5904stream}
\delta = 37.4026 + 0.2096 \alpha - 0.001578 \alpha^2
\end{equation}

\noindent Though M5 itself was not used in the fit, Equation 1 passes within
$4\arcmin$ of the cluster.

In Figure 3 we have conducted an identical analysis, but using a
coordinate system aligned with the stream in Figure 2. This coordinate
system has a pole at (R.A., dec) = ($326.626\arcdeg, +63.7782\arcdeg$),
and a zeropoint that places M5 at $\phi_1 = 0\arcdeg$. We have also attempted to remove the foreground contribution by fitting a 3rd order polynomial to the distribution after masking out M5 itself. Also shown is
the predicted orbit of M5 from Figure 1. While we do not generally
expect a tidal tail to precisely follow the orbit of its progenitor
\citep{eyre2011}, we note that the predicted orbit diverges somewhat from
the path of the stream, lying about $3\arcdeg$ north of the stream at
$\phi_1 \approx -50\arcdeg$. We discuss this further below.

The strongest portion of the stream extends to $\phi_1 \approx
-41\arcdeg$, whereupon the signal drops significantly to $\phi_1
\approx -50\arcdeg$. There may be an additional stream segment
extending from $\phi_1 \approx -62\arcdeg$ to $\phi_1 \approx
-85\arcdeg$. The significance of this segment is clearly marginal,
though its downward arc to the southwest is suggestively similar to
that of the predicted orbit. If we believe that this segment is real
then there appears to be a possible gap in the stream extending from
$\phi_1 = -50\arcdeg$ and $\phi_1 = -62\arcdeg$. Such a gap might be a
consequence of epicyclic motions of stars due M5's fairly eccentric
orbit, or it may have been caused by an encounter with with a massive
perturber, including perhaps a dark matter subhalo
\citep{carlberg2009,yoon2011}.

The lateral profile of the stream is shown in Figure 4. This profile
was measured using both $0.6\arcdeg$-wide and $2.6\arcdeg$-wide
rectangular masks extending from $\phi_1 = -47\arcdeg$ to $\phi_1 =
-10\arcdeg$. These masks were passed laterally across the
foreground-subtracted stream in Figure 3 and the total signal recorded
as a function of $\phi_2$.  The profile was then divided by the
standard deviation in the signal profile for all regions with
$|\Delta\phi_2| > 4\arcdeg$ to yield a measure of SNR (the T-index of
\citet{grillmair2009}). The strongest part of the stream is evidently
detected at $\sim 10\sigma$. Using the $0.6\arcdeg$-wide mask, we find
the full-width-at-half-maxium (FWHM) to be $1.7\arcdeg$.

How sensitive is our detection to the predicted proper motion profiles
in Figure 1? To test this, we offset the $\mu_{\alpha}\cos{\delta}$ and
$\mu_{\delta}$ profiles by specified amounts and measured the strength
of the stream as we did in Figure 4. Figure 5 shows the signal
strength as a function of proper motion offset. The signal clearly
drops quite quickly as we offset from the nominal profiles expected
for the orbit of M5. The widths of the peaks are consistent with the
proper motion uncertainties, which can exceed 1 mas/yr at our limiting
magnitude. Proper motion offsets in declination result in a more
precipitous decline in the signal strength than proper motion offsets
in $\mu_{\alpha}\cos{\delta}$. This appears to be in accord with Gaia's relative
proper motion uncertainties in the two directions in this part of the
sky. For stars with $18 < G < 20$ and satisfying our color-magnitude
constraints, the average $\mu_{\alpha}\cos{\delta}$ uncertainty is 0.74 mas/yr,
whereas for $\mu_{\delta}$ the average is 0.45 mas/yr.

We note in passing that the uncertainties in our M5 proper motion
profiles are dominated by the uncertainty in the distance to M5 (see
below). We are of course also subject to inaccuracies in our
adopted Galactic model. The uncertainties in the measured proper
motions and radial velocity for M5 \citep{gaia2018b} are so small as
to have essentially no visible effect on the profiles in Figure 1.

Based on $(i)$ the close alignment of the stream with M5, $(ii)$ the
similarity in distance and trajectory of the stream compared with that
of the model orbit, $(iii)$ that the stream is most strongly detected
using the proper motions predicted for M5's orbit, and $(iv)$ the fact
that the highest SNR is obtained using a filter based on the
color-magnitude distribution of stars in M5, we conclude that the
stream is most likely to be the trailing tidal tail of M5.

Our computed orbit for M5 predicts heliocentric distances that remain
relatively constant for much of the visible stream, rising gradually
to the west and peaking at at about 14.5 kpc at RA $\approx
135\arcdeg$. Thus, even if the stream extended around the Galaxy with
uniform surface density, we would expect the observed stream to fade
towards the west as fewer and fewer stars lie within Gaia's magnitude
limit. Moreover, the surface density of foreground stars with proper
motions similar to those of M5's orbit increases substantially as we
approach the Galactic plane to the west. This is shown in Figure
6, where we compare the predicted proper motions for M5's
orbit with the distribution of proper motions in the region of the
north Galactic pole.  Using a globular cluster luminosity function and
reducing the limiting magnitude by 1.5 magnitudes predicts that we
should see only one third as many stars. In addition, sampling Figure
5 or making a cut along $\phi_1$ just below the stream shows
that foreground contamination increases by more than a factor of
ten. Combining the reduction in stream stars with the increase in
background therefore predicts an SNR of $\approx 1$ at R.A. $=
135\arcdeg$. This is roughly consistent with the appearance
of the westenrmost segment. Figures 2 and 3 therefore do not rule out a stream
that continues westward with a constant surface density. Future deep,
multi-epoch surveys (e.g. WFIRST, \citet{sanderson2017}) may be able
to trace M5's tidal tail considerably further around the Galaxy.

There is no clear signature of M5's leading tail. While there appears
to be a pair of somewhat amorphous enhancements $\approx
10\arcdeg-15\arcdeg$ to the southeast of M5, these also coincide with
similarly shaped patches of significant reddening ((E - V) $\approx
0.4$, \citet{schlegel1998}). For the present, we attribute these
enhancements to inaccurate reddening corrections. M5's orbit predicts
that the leading tail should reappear south of the Galactic plane at
$\approx 11$ kpc, with distance increasing eastwards to over 20 kpc at
RA $> 320\arcdeg$. A cursory search for the leading tail south of the
Galactic plane did not yield any obvious candidate. Moreover, Gaia's
scan pattern in this region of the sky is almost parallel to M5's
predicted orbit, making an unambiguous identification somewhat
problematic. Based on the arguments above, we conclude that the SNR is
too low for detectability with existing Gaia measurements.

\subsection{Radial Velocity Measurements}

Matching by position the five hundred highest weighted stars in the
stream against the Sloan Digital Sky Survey Data Release 15
\citep{aguado2018}, we found a total of three matches. Of these, only
one star is classified as metal poor. This star (SDSS
J124931.41+195438.0) has (R.A., dec) =
($192.38091\arcdeg,19.910564\arcdeg$), $g, r, i = 19.46, 19.20, 19.1$,
[Fe/H] -1.5, and a heliocentric radial velocity of $-93 \pm 14$ km
s$^{-1}$. Figure 1 shows that the predicted velocity at this point
along M5's orbit is -73.5 km s$^{-1}$, in agreement at the
$1.4\sigma$ level. We therefore consider this a high probability
member of M5's trailing tail. 

As an aid for ongoing and future spectroscopic surveys, we provide
coordinates, Gaia magnitudes and colors, and proper motions for the
fifty highest-weighted candidate stream stars in Table 1. These
candidates span the strongest part of the stream ($188\arcdeg < $R.A.$
< 227\arcdeg$) and are ranked by relative weight, with the most
probable candidate listed first.

\section{Discussion} \label{discussion}

M5's trailing tail is somewhat broader than other globular cluster
streams. At a distance of 7.5 kpc, the measured FWHM of $1.7\arcdeg$
would correspond to $\approx 200$ pc. The stream also appears to be
less collimated and more ``wiggly'' than more populous streams like
GD-1 or Pal 5, though this may be partly a consequence of small number
statistics.

Another interesting feature of M5's tidal tail is that it does not
connect as directly or as strongly with the cluster as do, for
example, the tidal tails of Pal 5 \citet{odenkirchen2003} or NGC 5466
\citet{belokurov2006, grillmair2006}. The strongest part of the stream
is situated between $20\arcdeg$ and $30\arcdeg$ away from the cluster,
and there appears to be a gap or diminution between $5\arcdeg$ to
$10\arcdeg$ from the cluster itself. Does this argue against an
association between the stream and M5? As a simple experiment we set
the cluster to be 62 pc (the estimated tidal radius of a M5
\citet{harris1996}) and 200 pc (the width of the stream) further away
than its nominal distance and integrate new orbits around the
Galaxy. These integrations show that the stars in the trailing tail
should take between 3\% and 8\% longer to orbit the Galaxy than the
cluster itself. M5 is on a moderately eccentric orbit ($\epsilon
\approx 0.8$ \citep{gaia2018b}) and tidally stripped stars will
primarily be lost near perigalacticon. If the strongest part of the
stream in Figure 2 represents a pulse of stars lost during the last
perigalactic passage some $3 \times 10^8$ years ago, then we would now
expect the center of this pulse to be trailing the cluster by between
$14\arcdeg$ and $20\arcdeg$. The approximate agreement between the
observed distribution and the results of this experiment once again
point to an association between the stream and M5, and suggest that
the cluster is undergoing strongly episodic tidal stripping.

Could M5's trailing tail be detected using traditional,
photometry-only matched filter techniques applied to deeper
photometric surveys? We tested this by applying the matched-filter
technique described by \citet{grillmair2009} to the Pan-STARRs PS-1
catalog, using $g - r$ and $g - i$ color-magnitude diagrams for M5 to
a limiting magnitude of $g = 21.7$. This fainter magnitude limit would
nominally gain us between two and three times as many stream
stars. However, owing presumably to the lack of
proper motions and the concomitant severe contamination by
foreground stars, no trace of a stream could be detected. This
underscores the tremendous value of Gaia proper motions for detecting
very tenuous streams.

A possible issue with the identification of the stream with M5 is that
the stream stars fall on the ``wrong'' side of the M5's putative
orbit. Our adopted Galactic model predicts that we are very nearly in
the plane of M5's orbit, though very slightly ahead of this plane in
the direction of Galactic rotation. If this is indeed the trailing
tail of M5, then the stars would have been cast into slightly higher
Galactic orbits than that of M5. From our vantage point, this would
nominally put a trailing tail on or just slightly to the north of M5's
orbit. Yet the observed stream lies south of the putative orbit over
its entire length. This situation holds for a variety of orbits
computed by offsetting the velocities, proper motions, and parallax to
their $1\sigma$ limits (including a systematic error of 0.035 mas/yr
per \citet{gaia2018b}). The distance to M5 remains the most uncertain of
the input parameters, and the stream offset issue can be alleviated if
we set this to be 7 kpc. As shown in Figure 3, the predicted orbit
then lines up with the stream very well. However, such a distance
would depart by more than $2\sigma$ from the 7.6 kpc main-sequence
fitting estimate of \citet{sandquist1996}.

Alternatively, it is quite possible that our simple, spherical halo
model is inaccurate in this region of the Galaxy, and that either
non-sphericity and/or dynamical changes in the dark matter
distribution \citep{erkal2019, carlberg2019} are responsible for the
apparent divergence between orbit and stream. Realistic N-body
modeling of the stream may also shed light on whether the position of
the stream and M5's putative orbit can be plausibly
reconciled. Finally, despite our arguments above, it is possible
that the stream is not associated with M5. Spectroscopy of many
individual stars will be required to determine whether they are both
dynamically and chemically related to M5.

\section{Conclusions} \label{conclusions}

Using Gaia photometry and proper motion measurements we have detected
a stream of metal-poor stars that we believe to be the trailing tidal
tail of the globular cluster M5. The detection was made possible 
by the fact that the proper motions expected based on M5's orbit are
significantly different from those of most foreground stars in the
vicinity of the north Galactic pole. The most visible part of the tail
has an estimated surface density to $G = 20.0$ of 1.5 stars per square
degree, and a corresponding surface brightness of $> 35$ magnitudes
per square arcsecond. Owing to both increasing distance and foreground
contamination, the stream could plausibly extend much further around
the Galaxy but be largely undetectable in the Gaia data.

The predicted destruction rate of M5 is not particularly high among
globular clusters generally \citep{gnedin1997}, ranking somewhere in
the middle or near the bottom depending on the model used. The
detection of M5's trailing tail suggests that, given Gaia proper
motions, extended tidal tails for many other globular clusters may now
be detectable.  Previous work suggests that most if not all globular
clusters should have tidal tails and our detection demonstrates that
Gaia proper motion measurements may, in favorable circumstance,
provide a way of mapping the extended portions of these tails. Given
the relatively accurate measurements of the six dimensional phase
space coordinates now available for many of the nearby globular
clusters, finding such extended tidal tails will at a minimum aid in
our efforts to map the detailed contours and possible secular
evolution of the Galactic potential.

\acknowledgments

This work has made use of data from the European Space Agency (ESA) mission
{\it Gaia} (\url{https://www.cosmos.esa.int/gaia}), processed by the {\it Gaia}
Data Processing and Analysis Consortium (DPAC,
\url{https://www.cosmos.esa.int/web/gaia/dpac/consortium}). Funding for the DPAC
has been provided by national institutions, in particular the institutions
participating in the {\it Gaia} Multilateral Agreement.

The Pan-STARRS1 Surveys (PS1) have been made possible through
contributions by the Institute for Astronomy, the University of
Hawaii, the Pan-STARRS Project Office, the Max-Planck Society and its
participating institutes, the Max Planck Institute for Astronomy,
Heidelberg and the Max Planck Institute for Extraterrestrial Physics,
Garching, The Johns Hopkins University, Durham University, the
University of Edinburgh, the Queen's University Belfast, the
Harvard-Smithsonian Center for Astrophysics, the Las Cumbres
Observatory Global Telescope Network Incorporated, the National
Central University of Taiwan, the Space Telescope Science Institute,
and the National Aeronautics and Space Administration under Grant
No. NNX08AR22G issued through the Planetary Science Division of the
NASA Science Mission Directorate, the National Science Foundation
Grant No. AST-1238877, the University of Maryland, Eotvos Lorand
University (ELTE), and the Los Alamos National Laboratory.

Funding for the Sloan Digital Sky Survey (SDSS) has been provided by
the Alfred P. Sloan Foundation, the Participating Institutions, the
National Aeronautics and Space Administration, the National Science
Foundation, the U.S. Department of Energy, the Japanese
Monbukagakusho, and the Max Planck Society. The SDSS Web site is
http://www.sdss.org/.

The SDSS is managed by the Astrophysical Research Consortium (ARC) for
the Participating Institutions. The Participating Institutions are The
University of Chicago, Fermilab, the Institute for Advanced Study, the
Japan Participation Group, The Johns Hopkins University, Los Alamos
National Laboratory, the Max-Planck-Institute for Astronomy (MPIA),
the Max-Planck-Institute for Astrophysics (MPA), New Mexico State
University, University of Pittsburgh, Princeton University, the United
States Naval Observatory, and the University of Washington.

{\it Facilities:} \facility{Gaia, PS1, Sloan}

\clearpage

\begin{figure}
\epsscale{1.0}
\plotone{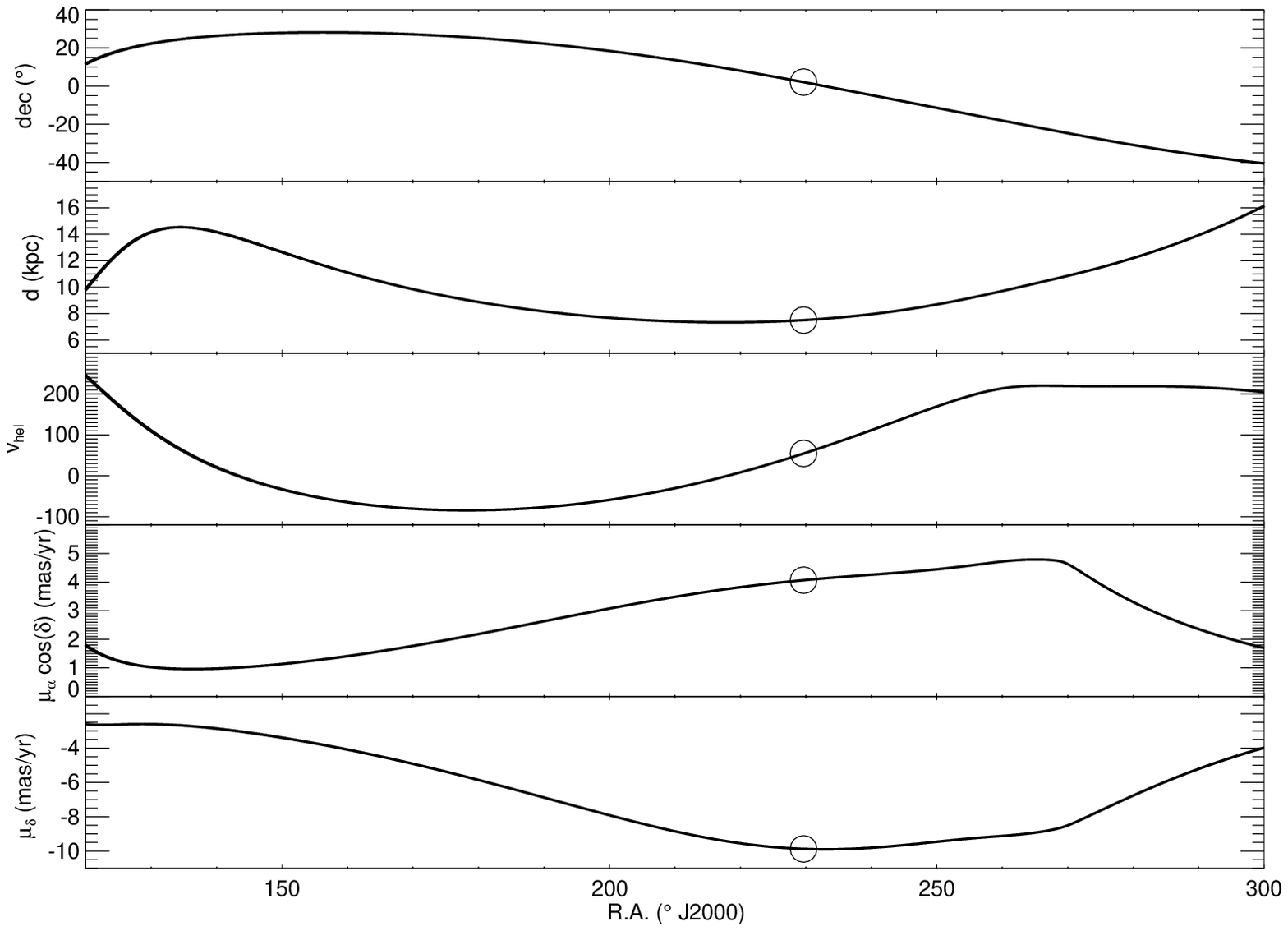}
\caption{The run of declination, distance, radial velocity, and proper
  motions with Right Ascension for the orbit of M5, as predicted using
  the Galactic model of \citet{allen1991}, updated with model parameters from
  \citet{irrgang2013}, an M5 distance of 7.5 kpc \citep{harris1996},
  and M5 proper motions from the \citet{gaia2018b}. The open circles show
  the values for M5 itself.}

\end{figure}

\begin{figure}
\epsscale{1.0}
\plotone{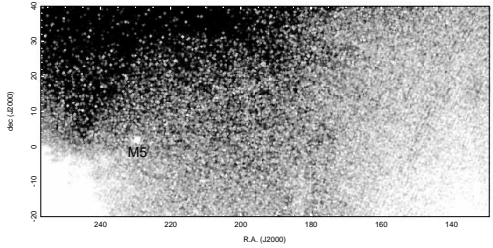}
\caption{Log stretch of a matched filter map of the region west of
  M5. Individual stars are weighted by their positions in the M5 $G,
  G_{BP} - G_{RP}$ color-magnitude diagram at a distance of 7.3 kpc
  (for the eastern two thirds of the region shown) and 12 kpc (for the
  westernmost third), and by their departure from
  the expected $\mu_\alpha\cos{\delta}$, $\mu_\delta$ profile of M5's orbit. The
  scale is $0.2\arcdeg$ per pixel and the map has been smoothed with a
  Gaussian kernel of $0.4\arcdeg$. The striations (most prominent on
  the right-hand side of the image) are a consequence of Gaia's scan
  pattern.}

\end{figure}

\begin{figure}
\epsscale{1.0}
\plotone{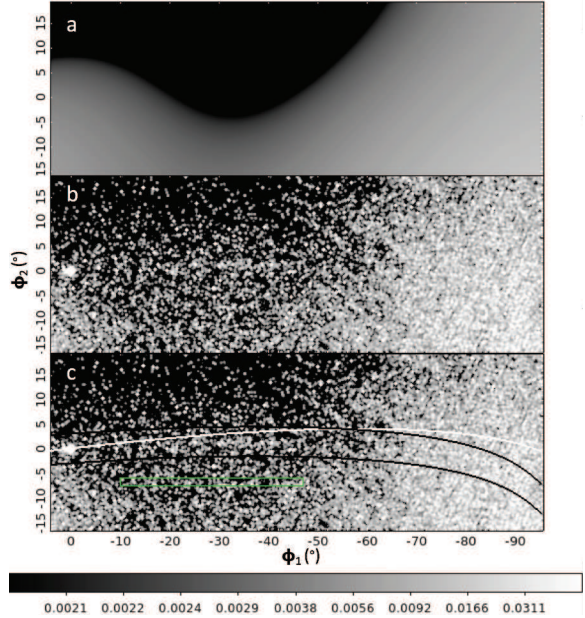}
\caption{As in Figure 2, but now using a coordinate system aligned
  with the stream (see text). $\phi_1$ and $\phi_2$ are now parallel and perpendicular to the stream, at least for the portion of the stream less than $40\arcdeg$ from M5. Panel a shows a 3rd order polynomial fit
  to the foreground distribution (after masking M5 itself) that is
  subtracted from the distribution to produce the maps shown in panels
  b and c. Individual stars are weighted so that a star lying
  precisely on the M5 color-magnitude locus as well as on the
  predicted $\mu_{\alpha}\cos{\delta}$ and $\mu_{\delta}$ profiles in Figure 1
  has a weight of unity. In panel c, the white curve shows the
  predicted orbit of M5, computed as described in the text. The black
  lines are $\pm 3\arcdeg$ offsets from an orbit computed assuming a
  distance for M5 of 7 kpc. The green box shows a mask used to
  measure the lateral profile shown in Figure 4.}

\end{figure}

\begin{figure}
\epsscale{1.0}
\plotone{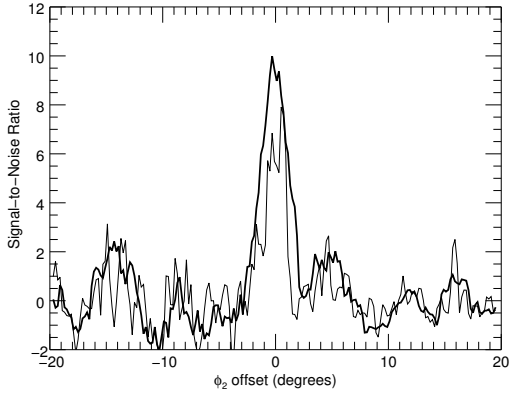}
\caption{The lateral profile of the stream. The thin black line shows
  the result of moving a $37\arcdeg$-long by $0.6\arcdeg$-wide
  mask across the stream in panel b of Figure 3, while the heavy black line shows
  the profile using a $2.6\arcdeg$-wide mask.}

\end{figure}

\begin{figure}
\epsscale{1.0}
\plotone{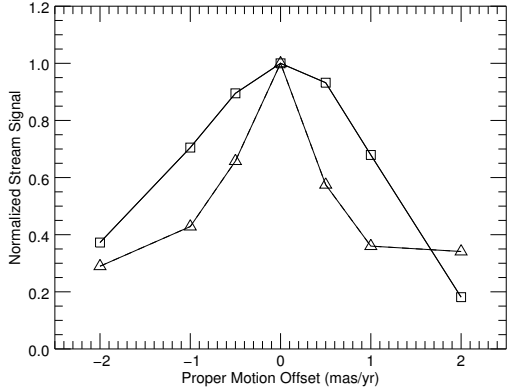}
\caption{The relative signal strength of the stream when constant
  offsets are applied to the proper motion profiles in Figure
  1. Squares show the results for offsets in $\mu_{\alpha}\cos{\delta}$ while
  triangles show the results when the $\mu_{\delta}$ profile is
  modified. The widths of the peaks are consistent with the relative
  uncertainties in the two directions.}

\end{figure}

\begin{figure}
\label{pmfig}
\epsscale{1.0}
\plotone{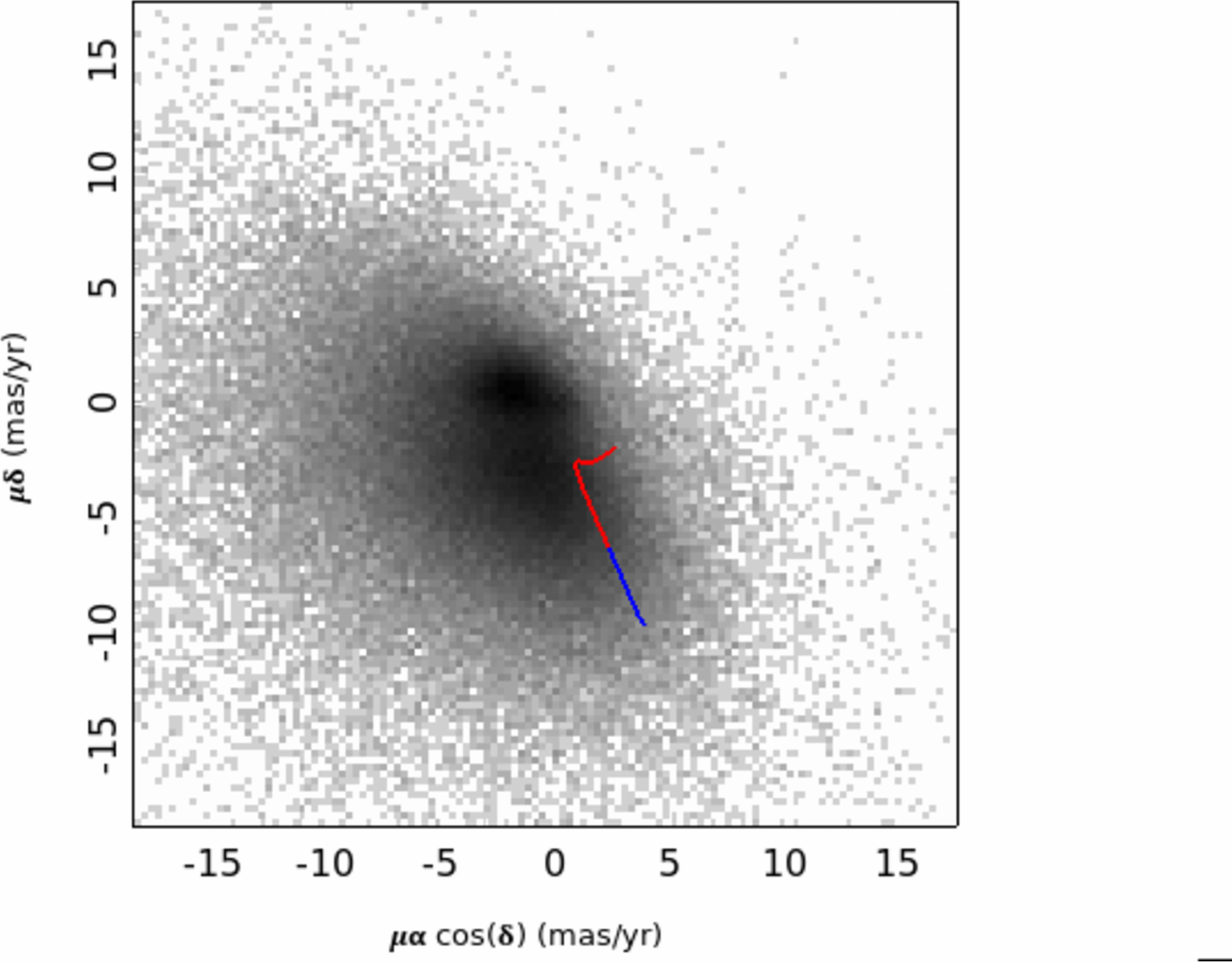}
\caption{The predicted proper motions for M5's orbit compared with the
  proper motion distribution of stars in the vicinity of the north
  Galactic pole. The blue curve shows the strongest, $50\arcdeg$ portion of the
  stream in Figure 2, while the red curve shows the portion of the
  orbit extending further to the west. The hook at the top of
  the predicted proper motion profile corresponds to the
  apogalacticon of the orbit.}

\end{figure}

\clearpage

%\floattable
\begin{deluxetable}{rccccccc}
%\rotate
\tabletypesize{\small}
\tablecaption{Candidate Stream Stars}
\tablecolumns{8}
%\tablewidth{0pt}

\tablehead{ \colhead{Rank} & \colhead{R.A. (J2015)} & \colhead{dec (J2015)} & \colhead{$G$} & \colhead{$G_{BP} - G_{RP}$} & \colhead{$\mu_\alpha \cos{\delta}$ (mas yr$^{-1}$)} & \colhead{$\mu_\delta$ (mas yr$^{-1}$)} & \colhead{Relative Weight}}

\startdata

\hline \\

   1 &   211.8487 &    10.5403 &  18.030 &   0.648 &    3.741 $\pm$  0.346 &   -9.117 $\pm$  0.096 & 1.000 \\
   2 &   214.1377 &    10.8701 &  18.624 &   0.661 &    3.439 $\pm$  0.518 &   -9.072 $\pm$  0.079 & 0.864 \\
   3 &   191.8278 &    19.3288 &  18.750 &   0.658 &    2.943 $\pm$  0.495 &   -7.059 $\pm$  0.117 & 0.836 \\
   4 &   221.2704 &     6.3224 &  18.349 &   0.633 &    3.604 $\pm$  0.453 &   -9.941 $\pm$  0.325 & 0.823 \\
   5 &   224.7915 &     5.2440 &  17.813 &   0.649 &    3.970 $\pm$  0.288 &   -9.758 $\pm$  0.357 & 0.690 \\
   6 &   219.6210 &     9.3874 &  18.481 &   0.626 &    3.470 $\pm$  0.454 &   -9.502 $\pm$  0.229 & 0.623 \\
   7 &   206.9149 &    13.8379 &  18.128 &   0.641 &    3.302 $\pm$  0.359 &   -8.887 $\pm$  0.304 & 0.612 \\
   8 &   189.1147 &    20.2214 &  18.553 &   0.631 &    2.236 $\pm$  0.429 &   -6.821 $\pm$  0.132 & 0.580 \\
   9 &   215.2481 &     9.1563 &  18.036 &   0.640 &    3.838 $\pm$  0.366 &   -9.582 $\pm$  0.484 & 0.565 \\
  10 &   200.5500 &    16.3797 &  18.725 &   0.660 &    2.524 $\pm$  0.525 &   -7.885 $\pm$  1.231 & 0.458 \\
  11 &   203.8473 &    13.6791 &  18.998 &   0.686 &    2.631 $\pm$  0.846 &   -8.046 $\pm$  0.453 & 0.458 \\
  12 &   225.6028 &     6.4901 &  18.418 &   0.612 &    4.303 $\pm$  0.394 &   -9.856 $\pm$  0.882 & 0.432 \\
  13 &   198.0241 &    17.3266 &  18.062 &   0.601 &    2.738 $\pm$  0.332 &   -7.762 $\pm$  0.355 & 0.420 \\
  14 &   202.0928 &    15.2032 &  17.942 &   0.650 &    3.382 $\pm$  0.382 &   -8.304 $\pm$  0.192 & 0.402 \\
  15 &   215.6956 &    10.3686 &  18.525 &   0.658 &    4.195 $\pm$  0.444 &   -8.972 $\pm$  0.182 & 0.373 \\
  16 &   214.1131 &     9.6728 &  18.593 &   0.676 &    3.546 $\pm$  0.486 &   -9.272 $\pm$  0.456 & 0.372 \\
  17 &   191.8149 &    16.7659 &  19.318 &   0.743 &    2.724 $\pm$  1.841 &   -7.245 $\pm$  0.394 & 0.371 \\
  18 &   194.1823 &    17.1066 &  18.529 &   0.642 &    3.553 $\pm$  0.521 &   -7.361 $\pm$  0.967 & 0.363 \\
  19 &   221.8236 &     7.5419 &  18.722 &   0.684 &    4.295 $\pm$  0.510 &   -9.537 $\pm$  0.122 & 0.362 \\
  20 &   199.0225 &    14.7108 &  19.222 &   0.704 &    3.293 $\pm$  0.799 &   -8.407 $\pm$  0.130 & 0.351 \\
  21 &   188.0709 &    17.4787 &  19.177 &   0.675 &    2.394 $\pm$  0.779 &   -6.701 $\pm$  0.122 & 0.337 \\
  22 &   216.2352 &     8.7810 &  19.415 &   0.734 &    3.002 $\pm$  0.882 &   -9.145 $\pm$  0.276 & 0.327 \\
  23 &   195.6849 &    15.8065 &  19.131 &   0.692 &    2.230 $\pm$  0.664 &   -7.666 $\pm$  0.667 & 0.324 \\
  24 &   210.9052 &    11.3475 &  19.595 &   0.781 &    3.522 $\pm$  0.863 &   -9.240 $\pm$  0.821 & 0.318 \\
  25 &   215.7512 &     6.6868 &  19.016 &   0.686 &    3.167 $\pm$  0.666 &   -9.635 $\pm$  0.036 & 0.293 \\
  26 &   194.0119 &    18.7484 &  18.648 &   0.695 &    3.016 $\pm$  0.468 &   -7.340 $\pm$  0.368 & 0.286 \\
  27 &   216.3491 &    10.2678 &  19.143 &   0.708 &    4.509 $\pm$  0.685 &   -8.981 $\pm$  0.621 & 0.283 \\
  28 &   201.8017 &    15.1407 &  19.615 &   0.765 &    3.949 $\pm$  1.281 &   -8.356 $\pm$  0.287 & 0.277 \\
  29 &   200.4833 &    16.7628 &  18.267 &   0.639 &    2.743 $\pm$  0.382 &   -7.539 $\pm$  0.577 & 0.276 \\
  30 &   215.8864 &     6.4038 &  18.756 &   0.565 &    3.355 $\pm$  0.771 &   -9.170 $\pm$  0.257 & 0.273 \\
  31 &   209.8048 &    10.4813 &  19.318 &   0.746 &    3.749 $\pm$  0.732 &   -8.350 $\pm$  0.082 & 0.272 \\
  32 &   210.2689 &    12.6871 &  19.811 &   0.811 &    3.111 $\pm$  1.208 &   -9.248 $\pm$  0.132 & 0.268 \\
  33 &   220.3115 &     7.1609 &  18.590 &   0.634 &    3.162 $\pm$  0.438 &   -9.549 $\pm$  0.071 & 0.266 \\
  34 &   217.4034 &     6.8449 &  18.993 &   0.695 &    2.829 $\pm$  0.645 &   -9.338 $\pm$  0.641 & 0.265 \\
  35 &   224.8610 &     4.9289 &  19.491 &   0.747 &    4.670 $\pm$  0.784 &   -9.577 $\pm$  0.869 & 0.262 \\
  36 &   197.7927 &    16.3254 &  18.502 &   0.631 &    3.473 $\pm$  0.393 &   -7.965 $\pm$  0.344 & 0.262 \\
  37 &   209.8688 &    11.0934 &  19.416 &   0.720 &    4.132 $\pm$  0.850 &   -8.683 $\pm$  0.053 & 0.261 \\
  38 &   209.1230 &    13.0422 &  19.078 &   0.703 &    4.395 $\pm$  0.741 &   -8.458 $\pm$  0.070 & 0.260 \\
  39 &   226.7971 &     6.8240 &  18.880 &   0.692 &    3.524 $\pm$  0.628 &  -10.534 $\pm$  0.047 & 0.259 \\
  40 &   205.6701 &    14.1893 &  19.461 &   0.735 &    3.359 $\pm$  1.023 &   -8.997 $\pm$  1.079 & 0.259 \\
  41 &   202.4636 &    15.4496 &  19.501 &   0.754 &    2.331 $\pm$  0.841 &   -7.971 $\pm$  0.446 & 0.257 \\
  42 &   216.7714 &     9.6120 &  18.744 &   0.680 &    4.442 $\pm$  0.687 &   -8.955 $\pm$  0.054 & 0.257 \\
  43 &   189.7340 &    20.4308 &  18.829 &   0.629 &    2.502 $\pm$  0.429 &   -6.752 $\pm$  0.191 & 0.252 \\
  44 &   218.7209 &     8.8706 &  19.974 &   0.850 &    4.217 $\pm$  1.579 &  -10.959 $\pm$  0.259 & 0.251 \\
  45 &   221.0086 &     6.6657 &  19.428 &   0.744 &    4.235 $\pm$  0.867 &  -10.441 $\pm$  1.050 & 0.245 \\
  46 &   212.5012 &    11.5745 &  19.523 &   0.776 &    3.038 $\pm$  0.962 &   -8.436 $\pm$  0.111 & 0.243 \\
  47 &   196.2040 &    17.9566 &  19.937 &   0.825 &    2.030 $\pm$  1.202 &   -6.985 $\pm$  0.051 & 0.242 \\
  48 &   196.3534 &    18.0855 &  19.751 &   0.797 &    2.016 $\pm$  0.994 &   -7.253 $\pm$  0.801 & 0.241 \\
  49 &   213.7135 &    10.4678 &  19.555 &   0.746 &    3.806 $\pm$  1.591 &   -8.274 $\pm$  0.128 & 0.241 \\
  50 &   199.6213 &    16.5163 &  19.149 &   0.676 &    2.654 $\pm$  0.731 &   -7.496 $\pm$  0.242 & 0.239 \\

\enddata
\end{deluxetable}

\end{document}